\documentclass[aps,prb,twocolumn,a4paper,superscriptaddress,floatfix,showpacs,preprintnumbers,amsmath,amssymb]{revtex4}
\usepackage[english]{babel}
\usepackage[utf8]{inputenc}
\usepackage{amsmath}
\usepackage{graphicx,epstopdf}
\usepackage{amssymb,epsfig,color,textcase}
\usepackage{hyperref}
\usepackage{doi}
\providecommand{\U}[1]{\protect\rule{.1in}{.1in}}

\def\feo2{FeO$_2$}
\def\feoh2{FeO$_2$H}

\def\etal{\textit{et al.}}

\begin{document}

\preprint{APS/123-QED}

\title{Hydrogenation driven formation of local magnetic moments in FeO$_2$H$_x$}

\author{Alexey O. Shorikov}
\email{shorikov@imp.uran.ru}
\affiliation{M.N. Miheev Institute of Metal Physics of Ural Branch of Russian Academy of Sciences - 620990 Yekaterinburg, Russia}
\affiliation{Department of theoretical physics and applied mathematics, Ural Federal University, Mira St. 19, 620002 Yekaterinburg, Russia}

\author{Alexander I. Poteryaev}%
\affiliation{M.N. Miheev Institute of Metal Physics of Ural Branch of Russian Academy of Sciences - 620990 Yekaterinburg, Russia}

\author {Vladimir I.~Anisimov}
\affiliation{M.N. Miheev Institute of Metal Physics of Ural Branch of Russian Academy of Sciences - 620990 Yekaterinburg, Russia}
\affiliation{Department of theoretical physics and applied mathematics, Ural Federal University, Mira St. 19, 620002 Yekaterinburg, Russia}

\author{Sergey V. Streltsov}%
\affiliation{M.N. Miheev Institute of Metal Physics of Ural Branch of Russian Academy of Sciences - 620990 Yekaterinburg, Russia}
\affiliation{Department of theoretical physics and applied mathematics, Ural Federal University, Mira St. 19, 620002 Yekaterinburg, Russia}

\date{\today}

\begin{abstract}
The electronic and magnetic properties of recently discovered new important constituent of the Earth's lower mantle \feoh2 were investigated by means of the density functional theory combined with the dynamical mean field theory (DFT+DMFT).
Addition of the hydrogen to the parent \feo2 compound, which is an uncorrelated bad metal, destroys the most important ingredient of its electronic structure - O-O molecular orbitals. In effect physical properties of \feo2 and \feoh2 turn to be completely different, \feoh2 is a correlated metal with a mass renormalization, $m^*/m \sim 1.7$, and magnetic moments on Fe ions become 
localized with the Curie-Weiss type of uniform magnetic susceptibility. 
\end{abstract}


\maketitle

\begin{figure}[hb!]
  \centering
  \includegraphics[width=0.20\textwidth]{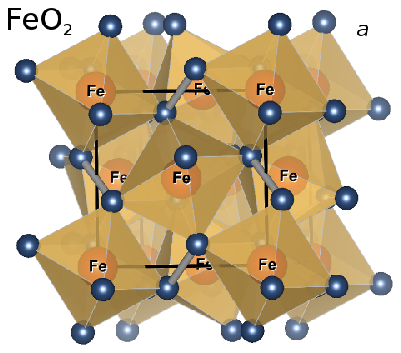}
  \;\;\;\;
  \includegraphics[width=0.20\textwidth]{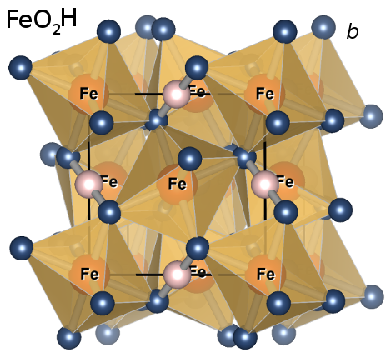}

  \vspace{3mm}

  \includegraphics [width=0.47\textwidth,clip=true]{fig3.eps}
  \caption {(Color online) Crystal structure for \feo2 ({\it a}) and \feoh2 ({\it b}) as obtained in the GGA relaxation at $P$=119~GPa.
  	         Fe, O, and H atoms are shown by orange, blue, and rose colors, respectively. Short O-O ``dimers'' that participate in forming antibonding $\sigma$ O-O molecular-orbitals are shown in gray.
  	         Corresponding total (black) and Fe 3$d$ (red) density of states for presented crystal structures are plotted below ({\it c} and {\it d}).}
  \label{fig:dft_dos}
\end{figure}

Iron oxides are the most important constituents of the Earth's mantle and core.
This is the reason why a lot of activity is concentrated on an investigation of their physical and chemical properties under high pressure.
Recent discovery of FeO$_2$ revised considerably this field~\cite{Hu2016}.
First of all, this compound was not known before 2016 and appears to be the most stable Fe oxide at pressure $>$100 GPa from the DFT point of view~\cite{Hu2016}.
Second, in contrast to other Fe oxides, which regarded as correlated materials with different type of transitions (metal-insulator, meta-magnetic, spin-state transitions, etc.),
Coulomb correlations were found to be almost unimportant in the iron dioxide~\cite{Streltsov2017}.
There are oxygen ``dimers''\footnote{We put quotas around dimers, since the distance in these pairs of oxygen atoms is still larger that in molecular oxygen.} in FeO$_2$ crystal structure, see Fig.~\ref{fig:dft_dos} (a), and molecular-orbitals due to these ``dimers'' determine electronic and magnetic properties of this compound.
These antibonding O-O molecular-orbitals appear exactly in the same energy region, where Fe $t_{2g}$ bands are located, hybridize with them and
this results in a formation of a pseudogap at the Fermi energy. In effect FeO$_2$ is an uncorrelated bad metal~\cite{Streltsov2017}.

However, in addition to pure FeO$_2$ there may exist hydrate FeO$_2$H at the Earth's lower mantle conditions~\cite{Nishi2017}.
Simple GGA+U calculations showed that FeO$_2$H is more stable than FeO$_2$ and H$_2$ separately~\cite{Nishi2017}.
The hydrogenation of FeO$_2$ was then approved by independent experiments at pressures 100--150~GPa~\cite{Hu2017} and resulting FeO$_2$H is now considered as one of the candidates, which forms a so-called $D''$ layer -- a core-mantle boundary~\cite{Liu2017,Mao2017}.

In spite of such a tremendous progress in study of the Earth's lower mantle, physical properties of FeO$_2$H remain mostly unexplored. In this Report we study electronic and magnetic properties of FeO$_2$H using calculations performed within the density functional (DFT) and dynamical mean-field (DMFT) theories and show that they are qualitatively different from the pure FeO$_2$. The hydrogenation makes FeO$_2$ a correlated material with local magnetic moments.

We start with simple DFT calculations, which give a fully relaxed (atomic positions, volume, and shape) crystal structure for an arbitrary pressure and allows one to study uncorrelated electronic structure of FeO$_2$H. The pseudo-potential VASP package~\cite{Kresse1996} and generalized gradient approximation (GGA)~\cite{Perdew1996} was utilized. Cutoff energy was set to 1000 eV,
$\mathbf{k}$-mesh consists of 343 points in the irreducible part of the Brillouin zone.
For most of the DFT-based calculations (if not stated specially) the pressure was chosen to be 119~GPa, that corresponds to known experimental crystal structure of FeO$_2$H~\cite{Nishi2017}.

Fig.~\ref{fig:dft_dos} shows crystal structures for \feo2 and \feoh2 and their corresponding density of states (DOS). In pure \feo2 octahedra are trigonally distorted with the same Fe-O distances, $d_{Fe-O}$=1.75~\AA~ (all the numbers in this and next paragraph correspond to the crystal structures obtained in the GGA calculations at $P=119$ GPa). 
%
%
The distance between two oxygen atoms forming ``dimers'' (shown by gray color in Fig.~\ref{fig:dft_dos}) is $d_{O-O}^{dim}=1.99$~\AA. This is much smaller than the length of O-O bonds forming edges of the FeO$_6$ octahedra ($d'_{O-O}$=2.60~\AA ~and $d''_{O-O}$=2.33~\AA), but still larger than the distance between oxygen ions in molecular oxygen, which is 1.21~\AA.
The presence of oxygen ``dimers'' leads to a formation of O-O molecular-orbitals and substantial modification of \feo2 electronic structure with respect to other iron oxides. If one imagines ``undimerized'' iron dioxide with the standard oxidation, an electron counting would give Fe$^{4+}$ and (O$_2$)$^{4-}$.
Similar to the case of pyrite, FeS$_2$, strong bonding-antibonding splitting in ligand-ligand ``dimers'' shifts antibonding $\sigma$ states upwards and reduces the oxidation of ligand's complex: 
(S$_2$)$^{2-}$ in iron disulfide and (O$_2$)$^{3-}$ in iron dioxide. The difference between pyrite and iron dioxide is in the strength of this bonding-antibonding splitting, which puts oxygen $\sigma-$antibonding bands exactly at the energy position of the Fe $t_{2g}$ bands, which results in further splitting and unusual valence of Fe: 3+\cite{Streltsov2017}.

The corner-shared octahedra of \feo2 are packed in such way that there are relatively large voids in between. These empty spaces are occupied by the hydrogen in case of \feoh2 and
it is crucial that H sits exactly in a middle of the oxygen ``dimers''. The most important structural consequences according to the GGA calculations are {\it (i)} approximately 10\% increase of the unit cell volume, from $V_{FeO_2}$=76.21~\AA$^3$ to $V_{FeO_2H}$=83.06~\AA$^3$, and {\it (ii)} increase of the distance in oxygen ``dimers'' to $d_{O-O}^{dim}=2.27$~\AA\footnote{Note that the GGA+U calculations give rather similar values: $d_{O-O}^{dim}=2.23$~\AA ~and $V_{FeO_2H}$=83.34~\AA$^3$.}. The Fe-O and O-O distances in the FeO$_6$ octahedra are nearly the same as in pure FeO$_2$: $d_{Fe-O}$=1.79~\AA, $d'_{O-O}$=2.69~\AA ~and $d''_{O-O}$=2.36~\AA.

Influence of the hydrogenation of \feo2 on the electronic structure is more dramatic.  It can be traced from the lower part of Fig.~\ref{fig:dft_dos}, where total and partial Fe 3$d$ DOS for both compounds (at the same pressure) are compared. The overall DOSes look very similar accounting for a band narrowing in case of FeO$_2$H. As we have pointed out this band narrowing comes from a huge volume enlargement: the unit cell volume of \feoh2 increases on $\sim$10\% with respect to \feo2. Hence, the Fe $t_{2g}$ bands shrink from 4.7~eV in \feo2 to 2.6~eV in \feoh2. 
The crystal field splitting between $t_{2g}$ and $e_g$ orbitals, that are centered around 3~eV,  is also affected by volume change and it is reduced from 4.07~eV to 3.46~eV~\footnote{In the octahedral environment, the $d$ level is split into triple degenerate $t_{2g}$ and double degenerate $e_g$ levels. An additional trigonal distortion leads to the splitting of  $t_{2g}$ 
level into double degenerate $e_g^{\pi}$ and $a_{1g}$ levels (above mentioned cubic $e_g$ states are named in this case as $e_g^{\sigma}$). The resulting splittings are
$\Delta_{e_g^{\sigma}-a_{1g}}^{FeO_2}$ = 3.97~eV, $\Delta_{a_{1g}-e_g^{\pi}}^{FeO_2}$ = 0.15~eV, and  
$\Delta_{e_g^{\sigma}-a_{1g}}^{FeO_2H}$ = 3.44~eV, $\Delta_{a_{1g}-e_g^{\pi}}^{FeO_2H}$ = 0.03~eV.
The fine details of the band structure are not important for our study, and thus, we keep using $t_{2g}$-$e_g$ notation over the text.}.
%
%
%
%
Nevertheless, this is not the most important change of the band structure. A careful checkup of the bands crossing the Fermi level shows that in the pure \feo2 there is a strong hybridization of the Fe $t_{2g}$ states with oxygen orbitals, which form $\sigma$-antibonding state in this energy region~\cite{Streltsov2017}
(the oxygen states can be seen in Fig.~\ref{fig:dft_dos} as a difference between total and Fe 3$d$ DOSes).
In case of \feoh2 these molecular orbitals are destroyed by the hydrogen, and therefore, the hybridization with oxygen in this energy range is extremely small. In effect the bands on the Fermi level are of pure $t_{2g}$ character as in many other iron oxides.

Thus, already on the DFT level one may argue that iron in FeO$_2$H behaves in a conventional way
(no molecular-orbitals and effects related to them) and should adopt ``3+'' valence state, as usual electron counting would suggest.
In FeO$_2$ the valence of iron is the same, ``3+'', but this is a consequence of a specific band structure, presence of a strong bonding-antibonding splitting as it was explained in Ref.~\onlinecite{Streltsov2017}. The fact that the Fe valence is the same in FeO$_2$ and FeO$_2$H is seen from nearly equal Fe-O bond distances obtained in the GGA calculations for these two compounds.

\begin {figure}[t!]
\centering
\includegraphics [width=0.47\textwidth,clip=true]{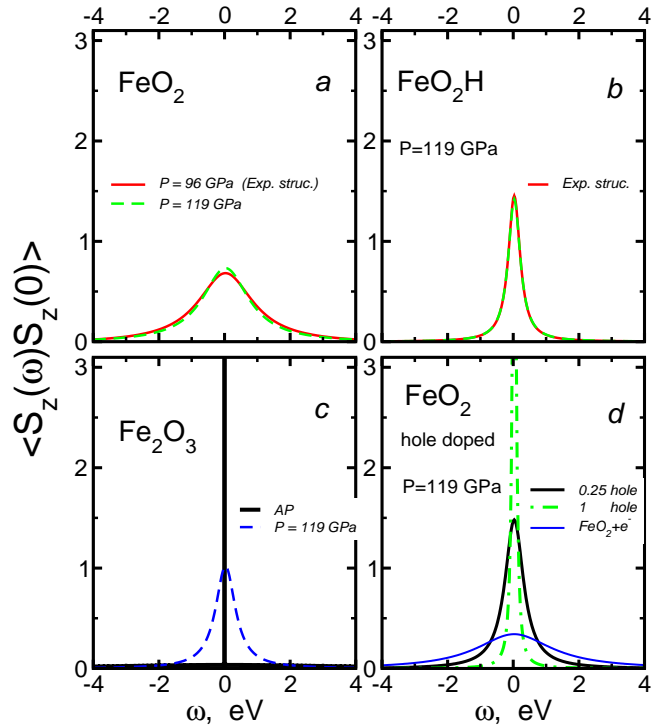}
\caption{(Color online) Local spin-spin correlation functions obtained within
	the DFT+DMFT formalism for FeO$_2$, FeO$_2$H, Fe$_2$O$_3$, and hole doped FeO$_2$ ($T=$1160~K). For details see legends and text. }
\label {fig:sts0_1}
\end {figure}

To proceed further with the magnetic properties investigation of compounds of interest we will use the DFT+DMFT method~\cite{Anisimov1997,Lichtenstein1997} as implemented in the AMULET code~\cite{AMULET}. This technique is very powerful at studying magnetic properties of materials in a paramagnetic state~\cite{Kotliar2006}. Another profit is to be able to make calculations for \feoh2, which is a correlated metal with localized magnetic moments, as we will show later. Thus, all compounds will be examined within the same framework regardless correlation strength.

In order to construct noninteracting GGA Hamiltonian, which included the Fe $3d$ and O $2p$ states, we use Quantum ESPRESSO~\cite{Giannozzi2009} and the Wannier function projection procedure~\cite{Korotin2008}.
The effective impurity problem was solved by the hybridization expansion (segment version) Continuous-Time Quantum Monte-Carlo method (CT-QMC)\cite{Gull2011}. 
To reduce off-diagonal elements of the hybridization function a transformation to a local coordinate system is performed by a diagonalization of the corresponding Fe 3$d$ blocks of the Hamiltonian, 
$\left[ \sum_{\vec k} H(\vec k) \right]_{Fe3d}$. In this case the off-diagonal elements of the hybridization function is less than 5 per cent its diagonal counterparts.
%
We used the same set of Coulomb parameters for all the structures and pressures under investigation, $U=6$ eV and $J_H=0.89$ eV ~\cite{Streltsov2017}. 
In order to avoid double counting of electron-electron interaction in the DFT+DMFT scheme, we use a self-consistent versions of fully localized limit (FLL)~\cite{Karolak2010} for the most of calculations. To benchmark correctness of our results with respect to a choice of double counting we carried out the calculations of the uniform magnetic susceptibility of \feoh2 using an around mean field  (AMF) correction~\cite{Karolak2010}.

Local spin-spin correlation functions,
$\langle \hat{S}_z(\omega) \hat{S}_z(0) \rangle$,
as obtained in the DFT+DMFT calculations, for different compounds are shown in Fig.~\ref{fig:sts0_1}.
$\hat{S}_z=\sum_m (\hat{n}_m^{\uparrow} - \hat{n}_m^{\downarrow})/2$, where
$\hat{n}_m^{\sigma}$ is an occupation operator for orbital $m$ and spin $\sigma$.
The width of this correlator is inverse proportional to the lifetime of spin moment. It is rather instructive to compare spin-spin correlation functions for FeO$_2$ and \feoh2  shown in Fig.~\ref{fig:sts0_1}a-b. The hydrogenation sharps peak and increases its value by factor of 2 approximately. Thus, one may see a dramatic increase of the spin localization in FeO$_2$H.

It is tempting to ascribe increase of the spin localization in FeO$_2$H to the volume enlargement. However, analysis of the volume dependence of the local spin-spin correlation function shows that this is not the case. One can see from Fig.~\ref{fig:sts0_1}a and Tab.~\ref{tbl:table} that there is only a minor change in the width of the correlator for FeO$_2$ going from $P=96$ to 119~GPa, while corresponding change of the volume is $\sim$9\%.
It should be noted here that the unit cell volume of \feoh2 at $P=119$ GPa is comparable with the volume of \feo2 at $P=96$ GPa.
This validates our assumption that change of volume plays a secondary role in explaining electronic and magnetic properties of \feo2 and \feoh2. Moreover, as one can see from Fig.~\ref{fig:sts0_1}d the electron doping going from \feo2 to \feoh2 leads to an opposite effect: decrease of the spin localization (see also discussion about different types of doping below). Thus, one might expect that the main reason of formation of the localized magnetic moments in FeO$_2$H is a destruction of the O-O ``dimers''. But how localized these moments are?

\begin{table}[tb!]
  \caption{\label{tbl:table} Unit cell volumes (third column) obtained at pressures shown in the second column
           by a full structural relaxation within the GGA method for various compounds (first column).
           In case of experimental structures the corresponding reference data were used.
           Instant squared magnetic moments calculated in the DFT+DMFT approach at $T=$1160~K are shown in the fourth column.}
  \begin{ruledtabular}
    \begin{tabular}{lccc}
	Compound			& P, GPa & V, \AA$^3$   & $\langle m_z^2 \rangle$, $\mu_B^2$ \\
    \hline
	\feo2 (exp.)\cite{Hu2016}	&   96   &   83.00	&   2.35	\\
	\feo2				&  119   &   76.21	&   2.45	\\
	\feo2 +0.25 hole			&  119   &   76.21	&   2.89	\\
	\feoh2 (exp.)\cite{Nishi2017}   &  119   &   83.03	&   2.26\\
	\feoh2				&  119   &   83.06	&   2.26	\\
	Fe$_2$O$_3$			&   AP   &  100.62	&  21.25	\\
	Fe$_2$O$_3$			&  119   &   67.60	&   2.19	\\
    \end{tabular}
  \end{ruledtabular}
\end{table}

In order to answer this question we compare FeO$_2$H with Fe$_2$O$_3$, where Fe is also 3+ (see Fig.~\ref{fig:sts0_1}c).
At ambient conditions this material is an insulator.
Fe$^{3+}$ ions are in the high-spin state with well developed local magnetic moments~\cite{Fujimori1986}.
This can be clearly seen from extremely sharp and strong peak in the correlation function and value of an instant squared magnetic moment, $\langle m_z^2 \rangle$ = 21.25~$\mu_B^2$, shown in Tab.~\ref{tbl:table}. Albeit  $R\bar{3}c$ phase of Fe$_2$O$_3$ used in the calculations does not exist above $P > 30$ GPa at $T \sim$1000~K, it is useful to study a degree of the spin localization in a hypothetical structure under pressures, where FeO$_2$ and FeO$_2$H can be formed. At $P=119$ GPa a volume of Fe$_2$O$_3$ decreases by 30\%,\footnote{Hypothetical $R\bar{3}c$ structure of Fe$_2$O$_3$ was obtained by full structural relaxation within the GGA.}  electronic bands become much broader that leads to a  metallicity, and as a result, to the broadening of spin-spin correlation function. The instant squared magnetic moment decreases one order in magnitude down to $\langle m_z^2 \rangle$ = 2.19~$\mu_B^2$ because of a transition from high-spin to low-spin state.  Comparing Fig.~\ref{fig:sts0_1} b and c one may see that the spins in FeO$_2$H turn out to be even more localized than in Fe$_2$O$_3$ at the same pressure. Certainly, the appearance of the hydrogen leads to the formation of the local magnetic moments in FeO$_2$H.

\begin {figure}[bt!]
  \centering
  \includegraphics [width=0.47\textwidth,clip=true]{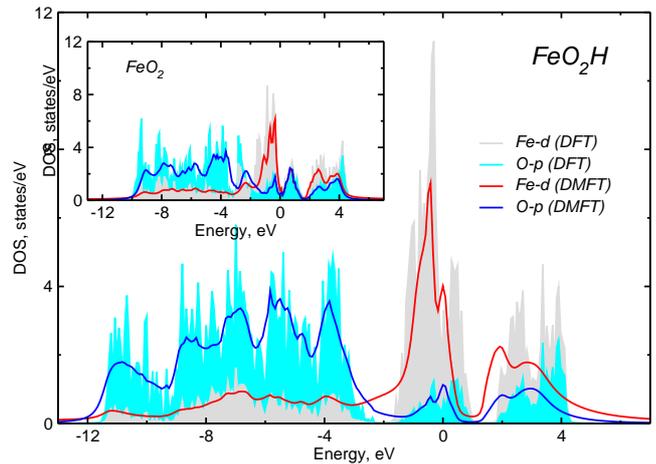}
  \caption {(Color online) Spectral functions for \feoh2 and \feo2 (inset). The DFT DOS are shown by filled gray (Fe) and cyan (O) colors.
  	         The DFT+DMFT spectral functions for $T=$1160~K are shown in red (Fe) and blue (O).}
  \label {fig:dmft_dos}
\end {figure}

It is worthwhile mentioning that the hydrogenation resembles hole doping of \feo2. Hydrogenating FeO$_2$ we add one electron to the system. Then the Fermi level should go to the right, crosses the pseudogap and then antibonding O-O band starts occupy (this would spins even less localized). In fact, both the DFT (Fig.~\ref{fig:dft_dos}) and DFT+DMFT (Fig.~\ref{fig:dmft_dos})  calculations demonstrate just an opposite behaviour: the Fermi level goes to the left and resides somewhere in the Fe $t_{2g}$ band. Adding hydrogen to FeO$_2$ we completely reconstruct electronic structure (break O-O molecular orbitals) and in some sense hydrogenation results in hole, not electron doping of FeO$_2$ (one may call it ``hole-doping-by-electron-doping''). Corresponding spin-spin correlation functions of hole doped FeO$_2$ and FeO$_2$H are indeed rather similar, see Fig.~\ref{fig:sts0_1} b and d. Thus, both the hydrogenation and hole doping of FeO$_2$ leads to a formation of localized magnetic moments.
\begin {figure}[tb!]
  \vspace{3mm}
  \centering
  \includegraphics [width=0.47\textwidth,clip=true]{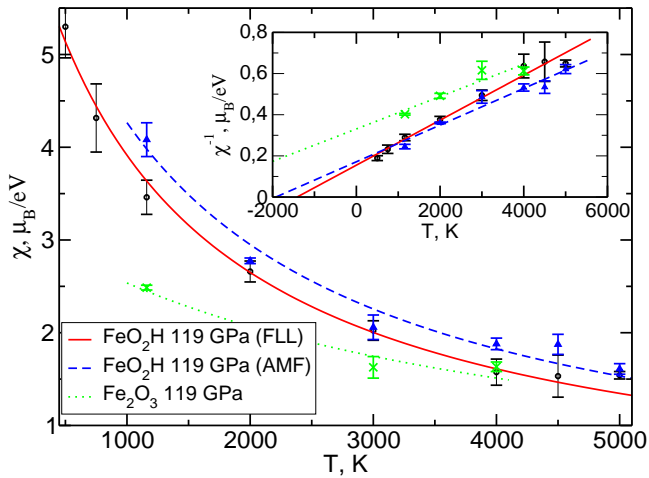}
  \caption {(Color online) Uniform magnetic susceptibility, $\chi(T)$, obtained by the DFT+DMFT method for FeO$_2$H (calculated using different types of the double counting, see text for details) and hypothetical Fe$_2$O$_3$ under the same pressure of 119 GPa. Inset shows an inverse $\chi(T)$.}
  \label {fig:chi}
\end {figure}

Uniform magnetic susceptibilities, $\chi(T)$, of \feoh2 and Fe$_2$O$_3$ are presented in Fig.~\ref{fig:chi}.
They strictly follow Curie-Weiss law typical for systems with localized magnetic moments. 
Our calculations of the uniform magnetic susceptibilities for \feoh2 show that the obtained results are robust to the choice of the double counting, with the estimated Curie-Weiss temperature to be $\Theta \sim-1500$~K ($\Theta \sim-1920$~K) for the FLL (AMF) scheme. This indicates substantial antiferromagnetic exchange interaction in FeO$_2$H.
$\chi(T)$ of \feoh2, calculated for different types of double counting, lie above its Fe$_2$O$_3$ counterpart, that confirms additionally the localized nature of magnetic moments in  \feoh2.
In contrast, the uniform magnetic susceptibility of FeO$_2$ grows with temperature~\cite{Streltsov2017}. The later behavior can be explained by the band structure peculiarities and FeO$_2$ should rather be considered as a material, which magnetic properties are described by band magnetism.

It has to be mentioned that while because of the large covalency\cite{Streltsov2017} there is no real difference in occupation numbers for FeO$_2$ and FeO$_2$H, both close six (6.2 electrons for FeO$_2$ and 5.7 electrons for FeO$_2$H), the influence of hydrogenation can be easily tracked down by investigating the DFT+DMFT spectral functions of \feo2 and \feoh2 shown in Fig.~\ref{fig:dmft_dos}.
We again start with FeO$_2$. The shape of the DFT+DMFT spectral function in FeO$_2$ remains almost unchanged with respect to DFT: the original DFT spectra become slightly smoothed by temperature and negligibly narrowed (see inset of Fig.~\ref{fig:dmft_dos} and Ref.~\onlinecite{Streltsov2017} for details).
This uncorrelated or band-like behavior comes from the fact that the Fermi level is in the pseudogap formed by the Fe $t_{2g}$ and mixture of Fe $t_{2g}$ and O-O antibonding states.  Hence, \feo2 is a bad metal with the band-type of magnetism.
\feoh2 demonstrates a completely different behavior. The effective mass enhancement, $m^*/m$, is 1.7 for $t_{2g}$  manifold and 1.3 for $e_g$ orbitals, which is comparable with values for classical Mott systems~\cite{Pavarini2004}.
Such a renormalization of the spectral weight leads to a quasiparticle peak narrowing in the vicinity of the Fermi level. 
One may argue that increased role of correlation effects is due to following factors: {\it i)} addition of hydrogen destroys oxygen ``dimers'' and effectively makes a hole doping of the Fe $t_{2g}$ subbands,
{\it ii)} reduced Fe $t_{2g}$ bandwidth increases $U/W$ ratio and moves \feoh2 to a more correlated regime.

Summarizing, we have studied the electronic and magnetic properties of \feoh2 by means of the DFT+DMFT method. We have found that hydrogenation changes drastically properties of the parent material. Hydrogen enlarges the volume of the unit cell by almost 10\% and, what is more important, destroys O-O ``dimers'' present in a pure FeO$_2$. In effect the Fermi level is moved from the pseudogap (in FeO$_2$) to the Fe $t_{2g}$ band, and FeO$_2$H turns out to be a correlated metal ($m^*/m \sim 1.7$) with a sharp quasiparticle peak at the Fermi level 
and well-formed local magnetic moments, while FeO$_2$ is bad uncorrelated metal, which magnetic properties can be described by itinerant theory of magnetism. The Fe ion adopts 3+ valency and is in the low-spin state ($3d^5$, $S=1/2$) at pressures of hundred GPa. Calculation of uniform magnetic susceptibility demonstrates that there is rather strong antiferromagnetic exchange coupling in FeO$_2$H (Curie-Weiss temperature $\Theta \sim -1500-2000$ K).

Our findings not only reveal a crucial role of hydrogenation on the physical properties of iron dioxide, but also cast doubt on possibility of consistent description of FeO$_2$ and FeO$_2$H in frameworks of the DFT. Neither GGA nor GGA+U approaches seem to be suitable for this, since while it may look like GGA+U is superior to GGA, because it partially takes into account Hubbard correlations, but in fact it breaks molecular-orbitals, which may lead to ``overstabilizion'' of FeO$_2$H with respect to FeO$_2$. This means that the use of more appropriate methods, like DFT+DMFT, may change previous results on structural and chemical stability of FeO$_2$H\cite{Nishi2017}.

S.S. is grateful to D. Khomskii for various useful discussions on physical properties of FeO$_2$. This work was supported by the grant of the Russian Scientific Foundation (project no. 14-22-00004).

\end{document}